\documentclass[preprint,showpacs,preprintnumbers,amsmath,amssymb, 
prd]{revtex4} 

\usepackage{graphicx} 
\usepackage{dcolumn} 
\usepackage{bm} 
\usepackage[ansinew]{inputenc} 
\usepackage[T1]{fontenc} 
\usepackage{euscript}

\begin{document} 

\preprint{} 

\title{\bf Conformal entropy from horizon states: 
Solodukhin's method for spherical, toroidal, and hyperbolic 
black holes in $D$-dimensional 
anti-de Sitter spacetimes} 

\author{Gonçalo A. S. Dias} 
%\email{gadias@fisica.ist.utl.pt} 
\author{José P. S. Lemos} 
%\email{lemos@fisica.ist.utl.pt} 
\affiliation{Centro Multidisciplinar de Astrofísica - CENTRA \\ 
Departamento de Física, Instituto Superior Técnico,\\ 
Universidade Técnica de Lisboa,\\ Av. Rovisco Pais 1, 1049-001 Lisboa, 
Portugal} 
\date{\today} 

\begin{abstract} 

A calculation of the entropy of static, electrically charged, 
black holes with spherical, toroidal, and hyperbolic compact and 
oriented horizons, in $D$ spacetime dimensions, is performed.  These 
black holes live in an anti-de Sitter spacetime, i.e., a spacetime 
with negative cosmological constant.  To find the entropy, the 
approach developed by Solodukhin is followed. The method consists in a 
redefinition of the variables in the metric, by considering the radial 
coordinate as a scalar field. Then one performs a $2+(D-2)$ 
dimensional reduction, where the $(D-2)$ dimensions are in the angular 
coordinates, obtaining a 2-dimensional effective scalar field 
theory.  This theory is a conformal theory in an infinitesimally small 
vicinity of the horizon. The corresponding conformal symmetry will 
then have conserved charges, associated with its infinitesimal 
conformal generators, which will generate a classical Poisson algebra 
of the Virasoro type. Shifting the charges and replacing Poisson 
brackets by commutators, one recovers the usual form of the Virasoro 
algebra, obtaining thus the level zero conserved charge eigenvalue 
$L_0$, and a nonzero central charge $c$. The entropy is then obtained 
via the Cardy formula. 
\end{abstract} 

%Meaning of PACS: 
%04.70.Dy: Quantum aspects of black-holes, evaporation, 
% thermodynamics 
%11.15.Hf: Conformal field theory, algebraic structures 
%04.60.-m: Quantum gravity 
%04.50.+h: Gravity in more than four dimensions, KK theory, 
% unified field theories, alternative theories of gravity 
%\keywords{Black Holes, Black Hole Entropy, Conformal Field Theory, 
%          Dimensional Reduction} 
%Use showkeys class option if keyword 

\maketitle 

\newcommand{\cequ}{\begin{eqnarray}} 
\newcommand{\fequ}{\end{eqnarray}} 
\newcommand{\beq}{\begin{equation}} 
\newcommand{\eeq}{\end{equation}} 
\newcommand{\anticomut}[2]{\left\{#1,#2\right\}} 
\newcommand{\comut}[2]{\left[#1,#2\right]} 
\newcommand{\comutd}[2]{\left[#1,#2\right]^{*}} 

\section{Introduction} 

The Bekenstein-Hawking formula 
for the entropy $S$ of a Schwarzschild 
black hole \cite{Bek,Haw}, given by 
\begin{equation} 
S=\frac{A}{4\,G}\,\left(\frac{k_{\rm B}\, c^3}{\hbar}\right)\,, 
\label{bkesteinhawkingformula1} 
\end{equation} 
being directly proportional to the 2-dimensional horizon 
area $A$ and inversely 
proportional to the constant of gravitation $G$, 
assigns to the black hole a gravito-geometric entropy. 
This entropy, together with the corresponding Hawking temperature 
$T=\hbar\,c^3/(8\,\pi\, k_{\rm B}\, G\,M)$ 
\cite{Haw}, where $M$ is the 
black hole's mass, has brought the black hole into the 
frame of thermodynamics (see, e.g. \cite{Wald book}). 
Thermodynamics is a phenomenological theory, which ignores in general 
the fundamental constituents, or degrees of freedom, of the 
object in consideration. On the other hand, statistical mechanics, 
together with the corresponding degrees of freedom, provides 
a much more fundamental way of understanding the phenomena. 
Thus, to have a better understanding of the 
degrees of freedom that give rise to this gravito-geometric 
entropy one should resort to statistical mechanics. 
Now, classically a black hole is characterized by its mass 
$M$, charge $Q$, and angular momentum $J$, less than a handful 
of  degrees of freedom. This characterizes the no-hair of black 
holes, i.e., the non existence of classical degrees of freedom. 
So where should one look for hairs? 
The dependence of the black hole entropy on $\hbar\,^{-1}$ 
in Eq. (\ref{bkesteinhawkingformula1}), 
suggests that the hairs, or the degrees of freedom, are essentially 
quantum mechanical. But what are they, and where are they located? 

These are problems that so far have no definite answers.  The black 
hole entropy could come from microscopic matter degrees of freedom 
induced by the quantum atmosphere of the matter fields generated by 
the Hawking radiation, in which case the degrees of freedom should be 
located in the horizon neighborhood, as has been advocated by several 
authors (see, e.g., \cite{rev} for a review). On the other hand, since 
the whole phenomenon is gravitational in nature, one can be led 
naturally to advocate the very existence of gravitational degrees of 
freedom that give rise to the gravito-geometric entropy $S$, given 
in Eq. (\ref{bkesteinhawkingformula1}). 
This idea was first raised by Bekenstein who speculated early that the 
degrees of freedom should be related to the quantization of the black 
hole area, emphasizing thus that they are in the 
gravitational field \cite{Bekensteingravitydof}. 

Theories that advocate to be quantum theories of gravity, such as 
string theory, loop quantum gravity, and induced gravity, have 
provided important calculations that give the entropy as in 
Eq. (\ref{bkesteinhawkingformula1}).  However, 
neither of the two questions raised above have a concrete answer. For 
instance, string theory, with its attached D-brane technology, has 
been successful in dealing with extreme black hole horizons, but not 
with non-extreme horizons or cosmological horizons (such as de 
Sitter).  Moreover, the celebrated Strominger and Vafa D-brane 
calculation \cite{Strominger and Vafa} is done in a regime for which 
the black hole is, in practical terms, a point like object from a 
distance, and not in a regime for which there is a black hole with a 
clear horizon, leaving one blind to the location of microstates. 

Thus, in order to try to understand the nature of the degrees of 
freedom one has to resort to other ideas, alternative to methods which 
try to find the entropy of a black hole from a fundamental theory. 
Following Bekenstein early idea \cite{Bekensteingravitydof}, 
Bekenstein and Mukhanov \cite{Bekenstein and Mukhanov} developed a 
heuristic scheme, where to each horizon Planck area one attaches a 
spin like degree of freedom, which provides a way to compute the number 
of accessible states, and thus the entropy, of a black hole. 

This idea, that the degrees of freedom are on the horizon, has been 
put into a precise mathematical formulation with the works of Carlip 
\cite{Carlip 1,Carlip 2,Carlip 3,Carlip 4,Carlip 5,Carlip 6}, which were 
developed and applied further by several authors, not only for general 
relativistic black holes \cite{Park Ho,Soloviev,Park Yee,Dreyer et 
al,Koga,Park,Kang}, but also for Lovelock and other theories with 
black holes \cite{Cvitan 1,Cvitan 2,Cvitan 3}, as well as in 
cosmological models with horizons (see, e.g, \cite{Brustein}). 
Essentially, Carlip's suggestion is that the hair comes from the 
symmetries of the horizon.  Now, the event horizon is not a true 
surface, and moreover one needs to know the global spacetime solution 
in order to trace its path. Therefore, the event horizon 
is not a candidate to extract 
information about the local micro degrees of freedom.  On the other 
hand, the apparent horizon has a local meaning and thus can be used as 
the boundary surface where to impose boundary conditions and search 
for symmetries.  The idea is that local quantum effects, manifested 
through Hawking radiation, break the diffeomorphism invariance of the 
spacetime and give sense to the prescribed boundary conditions in the 
vicinity of the apparent horizon.  For instance, imposing the 
conditions that the canonical generators are differentiable at the 
boundary horizon gives rise to a 2-dimensional conformal field 
theory with a non-vanishing central charge $c$. This, together with 
the level zero conserved charge eigenvalue, $L_0$, of the conformal 
theory gives, through Cardy formula \cite{Cardy}, which applies to 
2-dimensional conformal field theories in general, the entropy of the 
black hole. One one hand, this idea has arisen from the problem of 
universality, which is to explain why all types of approaches 
mentioned above agree with the semiclassical computation \cite{Carlip 
universality}. A possible answer is the one advanced above, 
that the result for the black 
hole entropy is 
controlled by a symmetry on the horizon inherited 
from the classical 
theory, which turns out to be a conformal symmetry 
hiding many underlying specific degrees of freedom 
\cite{Carlip 1,Carlip 2,Carlip 3,Carlip 4,Carlip 5}. 
On the other hand, Carlip's method has its 
physical basis on an observation by Strominger \cite{Strominger}, that 
by applying the Cardy formula \cite{Cardy} to the asymptotic symmetry 
group at infinity of the three-dimensional BTZ black hole spacetime 
\cite{BTZ}, one recovers the Bekenstein-Hawking expression. This 
followed the early result of Brown and Henneaux \cite{Brown Henneaux}, 
who showed that the asymptotic symmetry group at infinity of the 
three-dimensional anti-de Sitter spacetime is the conformal group in 
two dimensions with a non-vanishing central charge, and their 
additional remark that any quantum theory of such a type of spacetime 
should take this into account.  The three-dimensional BTZ black hole 
\cite{BTZ} belongs to that class of spacetimes, i.e., 
3-dimensional spacetimes asymptotically anti-de Sitter.  However, 
this calculation is heavily 
dependent upon the (2+1) dimensionality of the 
spacetime, (see \cite{Cadoni Mignemi} for an analysis along the same 
lines in (1+1) dimensions), 
and it locates the states responsible for the entropy at 
infinity, giving only an upper bound for the entropy of anti-de Sitter 
spacetime in three dimensions (which, in this case, it is just as 
good, since the maximum of entropy in a region arises by inserting a 
black hole in it).  The need to push these states onto the horizon 
resulted in the developments by Carlip \cite{Carlip 1,Carlip 2,Carlip 
3,Carlip 4,Carlip 5}. 

While putting the degrees of freedom on the horizon unequivocally, and 
in this way advancing that in order to understand the black hole 
entropy only the near horizon geometry of the black hole is relevant, 
Carlip's approach has not answered what those degrees of freedom are. 
Now, classically, black hole horizons are $(D-2)$-dimensional 
surfaces.  Quantum mechanically they can fluctuate, so it is plausible 
to expect that these fluctuations can be described by effective field 
theories inhabiting the two dimensions transverse to the horizon, 
essentially the time and the radial dimensions.  This idea has been 
put forward by Solodukhin \cite{Solodukhin}, who suggested that a 
possible candidate for the horizon states, in spherical symmetric 
black holes, is indeed the set of degrees of freedom of 
spherically symmetric fluctuations of the $(D-2)$-dimensional horizon 
sphere. More precisely, the statistical origin of black hole entropy 
is in the set of states of a 2-dimensional conformal field theory 
describing the fluctuations of the $(D-2)$-dimensional surface. Thus, 
while Carlip suggests that the hair is in the symmetries of the 
horizon, Solodukhin realizes it on radial fluctuations of the horizon. 
Solodukhin reduced then the problem of $D$-dimensional black holes to 
an effective 2-dimensional theory with fixed boundary condition on 
the horizon.  This theory, being conformal, admits then a chiral 
Virasoro algebra on the horizon, whose central charge $c$, and level 
zero operator eigenvalue $L_0$, yield directly, again through Cardy 
formula, 
the black hole entropy.  This near horizon conformal field 
theory approach by Solodukhin \cite{Solodukhin} has been refined in 
\cite{Carlip 7}, and extended to cosmological horizons like the de 
Sitter horizon \cite{Lin Wu}, as well as to black hole horizons 
in the Lovelock theory \cite{Cvitan et al Solodukhin}.  Similar ideas 
have been developed in \cite{Giacomini Pinamonti,Giacomini}. 

Now, Solodukhin's method relies on some particular assumptions, one of 
them being spherical symmetry, and all the black holes studied up to 
now have spherical horizons.  It is then worth understanding whether 
the topology of the horizon makes any difference to this method.  We 
investigate whether it works for black holes whose horizons have 
topologies other than spherical, 
such as in electrically charged black 
holes in asymptotically locally anti-de Sitter spacetimes in which the 
$(D-2)$ horizon sphere (i.e, a compact and orientable $(D-2)$ 
surface of constant 
positive curvature) 
is replaced by a  flat $(D-2)$ torus (i.e, a compact and 
orientable $(D-2)$ surface of zero curvature), or 
by a hyperbolic compact $(D-2)$ surface (i.e, a compact 
and orientable $(D-2)$ surface of constant negative curvature). 
In $D=4$ spacetime dimensions, 
the horizon can then be the usual 2-dimensional sphere 
(i.e., a Riemann surface with genus $g_{\rm enus}=0$), 
or a  2-dimensional flat torus (i.e., a Riemann surface with genus 
$g_{\rm enus}=1$) 
\cite{Lemos,Lemos Zanchin}, or a 2-dimensional 
hyperbolic torus, (i.e, a Riemann surface with genus 
$g_{\rm enus}\geq2$)
\cite{Smith Mann,Brill et al,Vanzo}. These 4-dimensional black 
hole solutions with compact orientable horizons have been 
generalized to $D$ dimensions in 
\cite{Birmingham} (see \cite{Santos Dias Lemos} for further 
references). More specifically, 
in this article we widen Solodukhin's method to include 
a generic set of static, electrically charged black holes, with 
negative cosmological constant, and with horizons with 
spherical, flat-toroidal, and hyperbolic-compact surface 
topologies. This generic set of black holes with different 
topologies is introduced in Sec. \ref{sec:blackholes}, 
where the corresponding 
metrics, entropies, and temperatures are given.  Note that, 
although the spherical case has been treated by Solodukhin 
\cite{Solodukhin}, we include it here because the three topologies can 
be treated in a unified way.  We then proceed in Sec. 
\ref{sectiondimensionalreduction} to a dimensional reduction, 
performing first a $2+(D-2)$ splitting, and then integrating in the 
$(D-2)$ angular 
coordinates, obtaining finally an effective 2-dimensional theory on the 
horizon with the corresponding field equation and 
constraints. In Sec. \ref{conformalentropy}, we show that there is 
conformal invariance on the horizon. Moreover, there exists an 
infinite number of conserved charges on the horizon, which provide a 
Virasoro algebra with non-vanishing central charge.  This analysis 
applies to all dimensions and the three topologies.  Using the Cardy 
formula we recover the Bekenstein-Hawking entropy.  In Sec. 
\ref{conclusions} we state the conclusions and final remarks. 

\section{The black holes} 
\label{sec:blackholes} 

\subsection{The action, and the metric and electric field} 
\label{sec:actionmetricelectricfield} 
The electrically charged anti-de Sitter black holes 
we are going to study are solutions of the 
Einstein-Maxwell action in $D-$dimensions, (we now put $\hbar=1$, 
$c=1$, and $k_{\rm B}=1$), 
\beq 
I = \int\,d^{D}x 
\sqrt{-g}\, \mathcal{L}\, \label{einsteinmaxwell1} 
\eeq 
with 
$g$ being the determinant of the metric $g_{\mu\nu}$, 
and the Lagrangian $\mathcal{L}$ being composed of 
the gravitational part and the electromagnetic 
part, and given by 
\beq 
\mathcal{L}=\frac{1}{16\pi G}\,(R-2\,\Lambda) 
-\frac14 F_{\alpha\beta}F^{\alpha\beta}\,, 
\label{einsteinmaxwell1.2} 
\eeq 
where $G$ is the $D$-dimensional Newton constant, 
$R$ is the Ricci scalar constructed from the 
Riemann tensor $R_{\alpha\beta\gamma\delta}$, 
$\Lambda$ is the cosmological constant, 
$F_{\alpha\beta}$ is the Maxwell tensor, and 
greek indices run the entire spacetime coordinates
$\alpha\,=\,0,\ldots,D-1$. 
Upon variation, the action 
(\ref{einsteinmaxwell1})-(\ref{einsteinmaxwell1.2}) gives the 
Einstein equation 
\begin{equation} 
R_{\alpha\beta}-\frac{1}{2}Rg_{\alpha\beta}+\Lambda 
g_{\alpha\beta}=8\pi\,G\, T_{\alpha\beta} \label{einsteinmaxwelleq} 
\end{equation} 
where $R_{\alpha\beta}$ is the Ricci tensor and 
$T_{\alpha\beta}=
\frac{1}{4\,\pi}\left( F_{\alpha\gamma}\,F_{\beta}^\gamma-\frac14\, 
g_{\alpha\beta} \, F_{\gamma\delta}\,F^{\gamma\delta}\right)$ is 
the electromagnetic energy-momentum tensor, and the 
Maxwell equation 
\begin{equation} 
\nabla_\alpha\,F^{\alpha\beta}=0\,. 
\label{maxwelleq} 
\end{equation} 

 From Eq. (\ref{einsteinmaxwelleq}) one can find a three-family of 
static, electrically charged black hole solutions with 
negative cosmological constant, 
with a parameter $k$, whose possible values are $k=1, 0, -1$. 
These three values correspond 
to black holes with spherical, flat-toroidal, and hyperbolic-toroidal 
topologies, respectively 
\cite{Lemos,Lemos Zanchin,Smith Mann,Brill et al,Vanzo,Birmingham,Santos 
Dias Lemos}. 
The horizons of these black holes are considered to be 
compact and orientable surfaces. 
The gravitational field of these black holes is described by 
the metric 
\beq 
ds^{2}=-f(r)\,dt^{2}+f(r)^{-1}\,dr^{2} 
+r^{2}\left(d\Omega^{k}_{D-2}\right)^{2} 
\,, 
\label{metric} 
\eeq 
where 
\beq 
f(r)=k-\frac{\Lambda}{3}r^{2}-\frac{2\,G\,M}{r^{D-3}}+ 
\frac{G\,Q^{2}}{r^{2(D-3)}} \,, 
\label{metricfunction} 
\eeq 
with $D\geq 4$, and $M$ and $Q$ are proportional to 
the ADM mass and electric charge, respectively. 
For instance, in the spherical case and for zero cosmological 
constant one has $M=\frac{8\pi \, m}{(D-2)\,\Sigma_{D-2}^{\,\,1}}$ and 
$Q^2= q^2 \left( \frac{8\pi }{(D-2)\,(D-3)} \right)$, 
where $m$ and 
$q$ are the ADM mass and electric charge, respectively, 
and $\Sigma_{D-2}^{\,\,1}$ is the area of the $(D-2)$ unit sphere 
\cite{MP}. 
For a generic dimension, and for generic types of black 
hole topologies, there is no closed expression for 
the ADM mass and electric charge, but for 
the matters we want to discuss here there is no need for 
such an expression. The angular part of (\ref{metric}) is given 
for each type of horizon, 
$k=1$, $k=0$, and $k=-1$, by 
\cequ 
\left(d\Omega^{\;\;1}_{D-2}\right)^{2}&=&d\theta_{1}^{2} 
+\sin^{2}\theta_{1}d\theta_{2}^{2}+ 
\ldots+\prod_{i=1}^{D-3}\sin^{2}\theta_{i}\,d\theta^{2}_{D-2}\,,\nonumber 
\\ 
\left(d\Omega^{\;\;0}_{D-2}\right)^{2}&=&d\theta_{1}^{2}+d\theta_{2}^{2}+ 
\ldots+d\theta_{D-2}^{2}\,,\nonumber \\ 
\left(d\Omega^{-1}_{D-2}\right)^{2}&=&d\theta_{1}^{2}+ 
\sinh^{2}\theta_{1}d\theta_{2}^{2}+\ldots  +\sinh^{2} 
\theta_{1}\prod_{i=2}^{D-3} 
\sin^{2}\theta_{i}\,d\theta^{2}_{D-2}\,, 
\label{angular metric} 
\fequ 
respectively. In the spherical case, $k=1$, one can 
take the range of coordinates as the usual one 
$0\leq\theta_1<\pi$, $0\leq\theta_i<2\,\pi$ for $i\geq2$. 
In the flat-torus case, $k=0$, the range is arbitrary, though finite 
for a compact surface, and we can choose 
$0\leq\theta_i<2\,\pi$. In the hyperbolic-compact surface  case, $k=-1$, 
the situation is in general more involved. For instance, 
in $D=4$ spacetime dimensions, 
the horizon is a 2-dimensional 
hyperbolic torus, i.e, a Riemann surface with genus 
$g_{\rm enus}\geq2$. 
In $D=5$ spacetime dimensions, 
the horizon is a 3-dimensional compact manifold ${\rm M}^3$ 
of constant negative curvature. In this case, the 
3-manifold ${\rm M}^3$ can be seen as a quotient space, 
${\rm M}^3={\tilde{\rm M}}^3/\Gamma$, where the universal covering 
${\tilde{\rm M}}^3$ 
is the hyperbolic 3-surface, and $\Gamma$ is a discrete 
subgroup of the isometry group of ${\rm M}^3$ \cite{Birmingham}. 
The range of coordinates in this case depends on 
$\Gamma$. 

The electric field of 
these solutions is radial and is given by 
\beq 
F_{ab}=F\,\epsilon_{ab} 
\label{electricfieldsolution} 
\eeq 
where 
$F$ is given by 
\beq 
F=-\frac12 \sqrt{ \frac{(D-3)(D-2)}{2\,\pi}}\frac{Q}{r^{D-2}}\,, 
\label{electricfieldsolutionF} 
\eeq 
and $\epsilon_{ab}$ 
is the antisymmetric tensor, with $a,b$ running through $0,1$, i.e., 
through the temporal $t$ and radial $r$ coordinates, respectively. 
We will split the greek indices $\alpha,\beta,\gamma,...$ into two
series of indices, the first denoted by lower case latin
$a,b,c,...=0,1$ for the temporal $t$ and radial $r$ coordinates,
respectively, and the second by upper case latin
$A,B,C,...=2,...,(D-1)$ for the corresponding angular coordinates.

\subsection{The entropy} 
\label{sec:entropy} 
The entropies associated with each topology all follow the area law. 
Indeed, according to \cite{Jac,Visser}, one has for 
black holes with bifurcating horizons, such as those of 
(\ref{metric})-(\ref{metricfunction}), the following expression, 
\beq 
S=-2\pi\oint_{\Omega} d^{D-2} y\sqrt{-h}\,\, Y^{\alpha\beta\gamma\delta}\hat
{\epsilon}_{\alpha\beta} 
\hat{\epsilon}_{\gamma\delta}\,, 
\label{entropiesingravitation0} 
\eeq 
with, 
\beq 
Y^{\alpha\beta\gamma\delta}\equiv \frac{\partial\mathcal{L}}{\partial R_
{\alpha\beta\gamma\delta}}\,. 
\label{defY} 
\eeq 
$\Omega$ is the $(D-2)$-dimensional horizon surface 
spanned by the coordinates $y^A$, $A=2,...,(D-1)$. 
Here $d^{D-2} y\sqrt{-h}$ is the induced measure of integration, where 
$h$ is the determinant of the 
induced metric $h_{AB}$ calculated on the horizon, with $h_{AB}$ 
being the metric in the $(D-2)$-dimensional 
spacetime, such that 
$h_{AB}\,dy^A\,dy^B= 
r^{2}\left(d\Omega^{k}_{D-2}\right)^{2}$, and 
we have split the greek indices $\alpha,\beta,\gamma,...$ into two
series of indices, the first denoted by lower case latin
$a,b,c,...=0,1$ for the temporal $t$ and radial $r$ coordinates,
respectively, and the second by upper case latin
$A,B,C,...=2,...,(D-1)$ for the corresponding angular coordinates.
$\mathcal{L}$ comes from Eq. (\ref{einsteinmaxwell1.2}). In addition, 
$R=R_{\alpha\beta\gamma\delta}g^{\alpha\gamma}g^{\beta\delta}=R_
{\alpha\beta\gamma\delta}\,\frac{1}{2}\, (g^{\alpha\gamma}g^{\beta\delta}- 
g^{\beta\gamma}g^{\alpha\delta})$, where the symmetries of the Riemann 
tensor 
$R_{\alpha\beta\gamma\delta}=R_{[\alpha\beta][\gamma\delta]}$ and $R_
{\alpha\beta\gamma\delta}=R_{\gamma\delta\alpha\beta}$ were considered 
($[\,]$ denoting antisymetrization), and 
$\hat{\epsilon}_{\alpha\beta}$ is the binormal to the bifurcate Killing 
horizon, normalized as $\hat{\epsilon}_{\alpha\beta}\hat{\epsilon}^
{\alpha\beta}=-2$. 
Now, following the definition (\ref{defY}), one has 
$Y^{\alpha\beta\gamma\delta}=\frac{1}{16\pi 
G}\,\left( \frac{1}{2}\, (g^{\alpha\gamma}g^{\beta\delta}-g^{\beta\gamma}g^
{\alpha\delta}) \right)$.  After 
rearrangements we get $S=\frac{1}{4G}\int_{\Omega}d^{D-2} y\sqrt{-h}$. 
Thus 
$\int_{\Omega}d^{D-2} y\sqrt{-h}=A_{D-2}$, 
where $A_{D-2}=\Sigma_{D-2}^{\,\,k}\,r_{\rm h}^{(D-2)}$ 
is the area of the $(D-2)$-dimensional horizon surface, 
$\Sigma_{D-2}^{\,\,k}$ 
being the area of the corresponding unit surface, with 
$k=1,0,-1$.
The radius $r_{\rm h}$ is the horizon radius 
determined by the equality 
$f(r_{\rm h})=0$. 
Thus, 
\beq 
S=\frac{A_{D-2}}{4\,G}\,, 
\label{entropiesingravitation1} 
\eeq 
in line with, e.g., the results in \cite{Brill et al} found through other 
methods for four dimensions. 
The entropy depends only on the Einstein term of the action, which has 
the same form for all of these solutions. In four dimensions, $D=4$, 
and for spherical symmetry, (\ref{entropiesingravitation1}) reduces to 
the Bekenstein-Hawking formula (\ref{bkesteinhawkingformula1}). 

\subsection{The temperature} 
\label{sec:temperature} 
Another important thermodynamic quantity for the black holes 
is the Hawking temperature $T$, the temperature at which the 
black holes emit black body radiation, which for these black holes 
is given by 
\beq 
T=\frac{1}{4\,\pi}\left[\frac{2\,(D-3)\,G\,M}{r_{\rm h}^{D-2}}- 
\frac{2\,(D-3)\,G\,Q^2}{r_{\rm h}^{2D-5}}-\frac{2\,\Lambda \, 
r_{\rm h}}{3}\right]\,. 
\label{hawkingtemperature1} 
\eeq 
This reduces to the Hawking temperature $T=1/(8\,\pi\,GM)$ 
for the 4-dimensional Schwarzschild black hole. 

\section{Dimensional reduction and effective 2-dimensional 
conformal field theory on the horizon} 
\label{sectiondimensionalreduction} 

\subsection{Preliminaries and motivation} 
\label{subsectionPreliminariesandmotivatiosection} 

Now, suppose we are dealing with a spacetime containing a 
black hole, which still preserving the original 
spherical, toroidal, or hyperbolic symmetry,  has a 
time dependence. This dependence could come from 
quantum vacuum fluctuations. 
Then a general class of $D-$dimensional metrics can be written as 
\beq 
ds^{2}=d\sigma^{2}+ d\Sigma^2\,, 
\label{2dformalismmetric} 
\eeq 
where 
\beq 
d\sigma^{2}= \gamma_{ab}(x)\,dx^{a}dx^{b} 
\,, 
\label{2Dmetric1} 
\eeq 
with $\gamma_{ab}(x)$ being a metric in a 2-dimensional 
spacetime with coordinates $x$, and 
where $x$ stands collectively for the time and spatial coordinates 
$x=(x^0,x^1)$, or, when convenient, for the null 
coordinates $x=(x^+,x^-)$, where $x^+=x^0+x^1$, 
and $x^-=x^0-x^1$, and 
\beq 
d\Sigma^{2}= h_{AB}(x,y)\,dy^A\,dy^B\equiv 
r^{2}(x)\left(d\Omega^{k}_{D-2}\right)^{2} 
\,, 
\label{2Dmetricangular} 
\eeq 
with $h_{AB}$ being the metric in the $(D-2)-$dimensional 
spacetime spanned by the coordinates $y$. 
The function $r(x)$ in Eq. (\ref{2Dmetricangular}) 
is the radial function, a type of dilaton function, 
in general dependent 
on the coordinates $x$. 
Note that the metrics (\ref{metric})-(\ref{angular metric}), 
being time independent, are a special case of this class of metrics. 
Indeed,  the 2-dimensional part of the 
metrics (\ref{metric})-(\ref{angular metric}) 
can be represented by the 2-dimensional metric 
$d\sigma^2=\gamma_{ab}(x)\,dx^{a}dx^{b}$, 
appearing in Eqs. (\ref{2dformalismmetric})-(\ref{2Dmetric1}), as 
\beq 
d\sigma^{2}= -f(x^1)\,{(dx^0)}^{2}+\frac{{(dx^1)}^{2}}{f(x^1)}\,, 
\label{2Dmetric2} 
\eeq 
with the function $f(x^1)$ vanishing at $x^1=x^1_{\rm h}$, 
where $x^1_{\rm h}$ is the 
horizon coordinate. In lightcone coordinates, one has 
$d\sigma^{2}= f \left(x_+,x_-\right) dx^+ dx^-$. 
For non-extremal black holes, one can write the function $f$ 
near the horizon as, 
\beq 
f(x^1)=\frac{4\,\pi}{\beta_{\rm{h}}}(x^1-x^1_{\rm{h}})+O 
\left((x^1-x^1_{\rm{h}})^{2}\right)\,, 
\label{f(x)} 
\eeq 
where $\beta_{\rm{h}}= 1/ (k_{\rm{B}}\, T)$ 
is the inverse temperature parameter, with $T$ 
given in Eq. (\ref{hawkingtemperature1}). 

Now, the event horizon of a black hole is not a true surface, but the 
apparent horizon $\mathcal{H}$ has a local meaning and thus can be 
used as the boundary surface where to impose asymptotic boundary 
conditions. The apparent horizon $\mathcal{H}$ for metrics 
(\ref{2dformalismmetric})-(\ref{2Dmetricangular}) 
 is defined as the surface $r(x)={\rm 
constant}$ whose outward normals are null, i.e., 
\beq 
\left.\gamma^{ab}\,\left(\nabla_{a}\,r\right) 
\left(\nabla_{b}\,r\right)\,\right|_{\mathcal{H}}=0\,, 
\label{hor} 
\eeq 
where $\nabla$ is the covariant derivative with respect to 
$\gamma_{ab}$.  This is a condition on the dilaton function $r(x)$. 
Note that condition (\ref{hor}) does not change under 
conformal transformations of the 2-dimensional metric 
$\gamma_{ab}$, i.e., transformations of 
the type $\gamma_{ab}\rightarrow {\rm e}^{2\rho}\, 
\gamma_{ab}$.  This led Solodukhin \cite{Solodukhin} to consider 
condition (\ref{hor}) as saying that the dynamics of the fluctuating 
metric is solely in the radial-dilaton function $r(x)$, 
while 
$\gamma_{ab}$ represents a class of conformal, but otherwise fixed, 
metrics.  It was further shown by Carlip \cite{Carlip 7} that 
the condition ``$\gamma_{ab}$ non-dynamical'' 
is consistent throughout. 
Interesting enough, Giacomini and Pinamonti 
\cite{Giacomini Pinamonti,Giacomini} have imposed 
otherwise, that $\gamma_{ab}$ is dynamical and the 
radial-dilaton function is non-dynamical, having found an effective, 
non-Lagrangian, theory which can be used to extract 
the horizon properties. 

Further arguments, which also apply to the spacetimes we are studying, 
led Solodukhin \cite{Solodukhin} to conclude that the generators 
$l_\xi\equiv \xi^a\,\partial_a$ of diffeomorphisms which preserve 
condition (\ref{hor}), and thus preserve the horizon $\mathcal{H}$, 
are conformal Killing vectors, satisfying the conformal Killing 
equation $\nabla_a\xi_b+\nabla_b\xi_a=\frac12 
\gamma_{ab}\nabla_c\xi^c$, generating thus the infinite dimensional 
group of conformal transformations on the horizon. Now, in general, 
the horizon $\mathcal{H}$ has several components, for instance, for 
the static black holes mentioned in Sec. \ref{sec:blackholes}, the 
horizon is composed of a future component $\mathcal{H}_{\rm f}$, 
defined in null coordinates by $x_-=0$ and $x_+>0$, and a past 
component $\mathcal{H}_{\rm p}$, defined by $x_+=0$ and $x_-<0$. In 
the static case these components intersect.  We will be interested in 
the future horizon $\mathcal{H}_{\rm f}$.  In this case, the future 
generators of $\mathcal{H}_{\rm f}$ are given by 
$l_\xi=\xi^+(x^+)\,\partial_+$. In turn, its Fourier components 
$l_n={\rm e}^{i\,n\,x^+}\,\partial_+$ form a copy of the Virasoro 
algebra, i.e., of $[l_n,l_m]=i\,(m-n)\,l_{n+m}$, with respect to the Lie 
bracket $[\xi_1,\xi_2] =(\xi_1\,\xi_2'-\xi_2\,\xi_1')$ 
\cite{Solodukhin}. 

These arguments show unequivocally that the horizon has some conformal 
structure built in it, and leads one to seek an effective field theory 
near the horizon of a black hole.  Solodukhin \cite{Solodukhin} worked 
out in detail the case of spherically symmetric general relativity and 
showed that treating the horizon as a boundary leads to a theory, that 
is indeed a conformal theory possessing the Virasoro algebra of the 
generators of the horizon, plus a central extension, 
with a central 
charge which is a crucial term for finding the entropy of the black 
hole.  We will work out in detail black holes with toroidal and 
hyperbolic topology in general relativity (including 
for completeness the spherical case first treated by 
Solodukhin \cite{Solodukhin}) 
and show that all the different topologies 
yield a conformal field theory at the horizon whose extended 
Virasoro algebra possesses also a central term. 

\subsection{Dimensional reduction and effective 2-dimensional 
conformal theory } 
\label{subsectionDimensionalreductionandeffectivetheory} 

In order to find an effective 2-dimensional field theory, 
we now apply dimensional reduction to the action 
(\ref{einsteinmaxwell1}) using the $2+(D-2)$ splitting 
used in Eq. 
(\ref{2dformalismmetric}) for the metrics 
(\ref{metric})-(\ref{angular metric}). 
Since we have agreed that the dynamical field is the radial coordinate 
$r$, this will be kept in the action. 
Then, one first dimensionally reduces the gravitational 
part of the action (\ref{einsteinmaxwell1}) 
to a 2-dimensional spacetime represented by (\ref{2Dmetric1}). 
In addition, one also has to reduce the electromagnetic part 
of the action. 
Dimensionally reducing the action 
(\ref{einsteinmaxwell1}) gives then 
\begin{eqnarray} 
I = -\frac{\Sigma_{D-2}^{\,\,k}}{16\pi G} 
\int\,d^{2}x &&\sqrt{-\gamma}\, 
\left[r^{D-2}R+(D-3)(D-2)r^{D-4}\left(\nabla r 
\right)^2+\right.\nonumber\\ 
&&\left. 
k(D-3)(D-2)r^{D-4} 
-2\Lambda\,r^{D-2} +8\,\pi\,G F^{2}r^{D-2}\right], 
\label{action2D1} 
\end{eqnarray} 
where $R$ is now the scalar curvature in two dimensions, 
$\gamma_{ab}$ is the 2-dimensional metric, $\Lambda$ is still 
the cosmological 
constant, $\Sigma_{D-2}^{\,\,k}$ 
is the area of the corresponding unitary surface, 
a factor that comes from the integration of the angular coordinates, 
and $F$ is given in Equation (\ref{electricfieldsolutionF}). 
For the spherical case, $k=1$, one gets 
$\Sigma_{D-2}^{\,\,1}={2\pi^{\frac{D-1}{2}}}/\,{\Gamma 
\left(\frac{D-1}{2}\right)}\,$. 
For the flat-torus case, $k=0$, choosing 
the range of angular coordinates between 0 and $2\,\pi$, 
gives $\Sigma_{D-2}^{\,\,0}=\left(2\pi\right)^{D-2}$. 
For the hyperbolic-compact surface  case, $k=-1$, 
the situation is more complicated. For a 2-dimensional 
compact hyperbolic horizon (in the spacetime with $D=4$), one has, 
by the Gauss-Bonnet theorem, 
$\Sigma_{\,\,2}^{-1}=4\,\pi\,(g_{\rm enus}-1)$, where $g_{\rm enus}$ 
is the genus of 
the Riemann surface, with $g_{\rm enus}\geq2$. For higher horizon 
dimensions, the area of the unitary compact surface 
depends on the isometry group one considers 
acting on the hyperbolic space \cite{Birmingham}. 

Now redefine the coordinate $r$ as the following dilaton field $\phi$, 
\cequ 
\phi&=& \frac{1}{2\,q}\, \left(\frac{D-2}{D-3}\right)\, 
C^{\frac12}\,\left(\frac{r^2}{r_{\rm h}}\right)^{\frac{D-2}{2}}, 
\label{redef1} 
\fequ 
where $q$ is a constant to be determined later, 
and $C=\frac{\Sigma_{D-2}^{\,\,k}}{2\pi G} 
\left(\frac{D-3}{D-2}\right)$, and perform a conformal 
transformation of the 2-dimensional metric $\gamma_{ab}$, 
into a metric  $\bar{\gamma}_{ab}$, 
\cequ 
\gamma_{ab}&=& 
\left(\frac{\phi_{\rm h}}{\phi}\right)^{\frac{D-3}{D-2}} 
{\rm e}^{\frac{1}{q^2\left(\frac{D-3}{D-2}\right)} 
\,\frac{\phi}{\phi_{\rm h}}} 
\;\,\bar{\gamma}_{ab}\,, 
\label{redef3} 
\fequ 
where 
$\phi_{\rm h}=\frac{1}{2q} 
C^{\frac12}\left(\frac{D-2}{D-3}\right)(r_{\rm h})^{\frac{D-2}{2}}$ is the 
classical value of the field on the horizon. Then 
the action (\ref{action2D1}) becomes a Liouville type action, 
\beq 
I=-\int\, d^{2}x\sqrt{-\bar{\gamma}} 
\left[\frac{1}{2}(\bar{\nabla} \phi)^{2}+ 
\frac{1}{2}q^2\left(\frac{D-3}{D-2}\right)\phi_{\rm h}\,\phi \bar{R} 
+ U(\phi)\right]\,, 
\label{actionfinalD>4} 
\eeq 
where $\bar{\nabla}$ is the derivative associated with the new metric 
$\bar{\gamma}_{ab}$, 
$\bar{R}$ is the scalar curvature associated with $\bar{\gamma}$, 
derived from the original scalar $R$ through 
$\bar{R}=\Omega^{-2}\left[R-2(D-1)\gamma^{ab}\nabla_a\nabla_b(\ln \Omega)- 
(D-2)(D-1)\gamma^{ab}(\nabla_a\ln\Omega)(\nabla_b\ln\Omega)\right]$, 
with $\Omega=\left(\frac{\phi}{\phi_{\rm h}}\right)^{\frac{D-3}{2\,(D-2)}} 
\exp\left(-\frac{\phi}{2\,q^2\left(\frac{D-3}{D-2}\right)\phi_{\rm 
h}}\right)$ 
(cf., e. g., \cite{Wald book}), and where the potential $U(\phi)$ is 
given by 
\cequ 
U(\phi)=\left(\frac{\phi_{h}}{\phi}\right)^{\left(\frac{D-3}{D-2}\right)} 
e^{\frac{\phi}{q^2\left(\frac{D-3}{D-2}\right)\phi_{\rm h}}} 
&&\left[\frac{k}{2}(D-2)^{2} 
C^{\frac{2}{D-2}}\left(\frac{D-3}{D-2}\right)^2 
q^2\phi_{\rm h}\phi\right.\nonumber\\ 
&&\left.+q^2\left(\frac{D-3}{D-2}\right)\phi_{\rm h} \, 
\phi\,(4\,\pi\,G\,F^{2}-\Lambda) \right]\,, 
\label{potentialD>4} 
\fequ 
Except for the metric $\bar{\gamma}_{ab}$, 
in the subsequent expressions the bar in the several other 
quantities will be dropped for simplicity of notation. 

The action (\ref{actionfinalD>4}) provides an effective theory 
for the dilaton field $\phi$, which in turn represents the 
oscillations of the radial coordinate near the horizon. 
To find the equations of these oscillation one has to 
vary the action (\ref{actionfinalD>4}). 
Its variation with respect 
to the dilaton field gives the following equation of motion for 
$\phi$ itself, 
\beq 
\square\,\phi=\frac{1}{2}\,q^2\,\left(\frac{D-3}{D-2}\right)\,\phi_{\rm h} 
\, 
R+U'(\phi)\,, 
\label{equ:phi} 
\eeq 
where $\square$ is the d'Alembertian $\square=\partial_{a}\,\partial^{a}$. 
The  variation of the action (\ref{actionfinalD>4}) 
with respect to the metric $\bar{\gamma}_{ab}$ 
gives $\delta\,I=\frac12\int d^2x\, T^{ab}\,\delta\bar{\gamma}_{ab}=0$, 
i.e., 
\cequ 
T_{ab} 
&\equiv&\frac{1}{2}\partial_{a}\phi\partial_{b}\phi- 
\frac{1}{4}\bar{\gamma}_{ab}(\nabla\phi)^{2}+\frac{1}{2}q^2 
\,\left(\frac{D-3}{D-2}\right)\,\phi_{\rm h}(\bar{\gamma}_{ab} 
\square\phi-\nabla_{a}\nabla_{b}\,\phi)\nonumber\\ 
&&-\frac{1}{2}\bar{\gamma}_{ab}U(\phi)=0\,, 
\label{energymomentumtensor} 
\fequ 
where we have used that the Einstein 
tensor $G_{ab}=R_{ab}-\frac12\bar{\gamma}_{ab} R$ is identically zero 
in two dimensions. This equation is a constraint equation. 

So, now one has a 2-dimensional theory, with constraints, 
for a scalar field $\phi$, strikingly resembling bosonic 
string theory \cite{GreenSchwarzWitten}. 
However, there is a snag; as it is, this 
theory is not conformally invariant everywhere. 
Indeed, 
the trace of $T_{ab}$, $T^a_a=\bar{\gamma}^{ab}T_{ab}$, is given by 
\beq 
T^a_a=\frac{1}{2}\,q^2 
\,\left(\frac{D-3}{D-2}\right) 
\phi_{\rm h}\,\square\,\phi-U(\phi)\,, 
\label{traceenergymomentumtensor} 
\eeq 
and is not equal to zero in general. 
Now, under conformal transformations, 
$\bar{\gamma}_{ab}\rightarrow \bar{\gamma}_{ab}(x)+\lambda(x) 
\bar{\gamma}_{ab}(x)$, one has $\delta \bar{\gamma}_{ab} 
=\lambda(x)\bar{\gamma}_{ab}(x)$, and so 
the action varies as 
$\delta\,I=\frac12\int d^2x\,\lambda(x)\,T^a_a\,$. 
Thus, one sees that the vanishing of the trace of the 
energy-momentum tensor is a necessary condition for invariance of the 
action 
under conformal transformations. 
This means that the 2-dimensional theory for the scalar $\phi$ is 
not, in general, a conformal theory. Interestingly enough, 
it becomes conformal in an infinitesimally small vicinity of the horizon. 

To see this, choose, close to the horizon,  the following 
coordinate $z$ 
\beq 
z=\int^{x^1}\frac{dx}{f(x)}= 
\frac{\beta_{\rm h}}{4\,\pi}\ln(x^1-x^1_{\rm h})\,, 
\eeq 
where $f(x)$ is given by Eq. (\ref{f(x)}). 
Note that this asymptotic region, close to the horizon, 
is a region of  negative high values of $z$, $z\rightarrow-\infty$. 
The metric function $f(x)$ given in Eq. (\ref{f(x)}) becomes, 
\beq 
f(z)=f_{0}\,e^{\frac{4\,\pi}{\beta_{\rm{h}}}\,z}\,, 
\label{f(z)} 
\eeq 
where $f_0$ is some constant. 
Thus $f(z)$ vanishes exponentially fast at the horizon. 
We can see that after this transformation the metric has gone from 
$\bar{\gamma}_{ab}$ 
to a metric of the form $f(z)\, \eta_{ab}$, 
where $\eta_{ab}$ is the Minkowski metric in two dimensions. 
We also put $x^0=t$, so that the 2-dimensional 
plane is now defined by the coordinates $(t,z)$. 
Then, Eq. (\ref{equ:phi}) is, in coordinates $(t,z)$, 
given by 
\beq 
-\partial^{2}_{t}\phi+\partial^{2}_{z}\phi=\frac{1}{2} 
\, \left(\frac{D-3}{D-2}\right)\, q^2 
\phi_{\rm h}\,R\,f(z)+f(z)\,U'(\phi)\,. 
\label{eq1} 
\eeq 
At the horizon, for 
large negative values of $z$, the right 
hand side of Eq. (\ref{eq1}) 
vanishes exponentially fast. It then becomes, at $z\rightarrow-\infty$, 
\beq 
\partial^{2}_{t}\phi-\partial^{2}_{z}\phi=0\,, 
\label{dalemberteq} 
\eeq 
a simple d'Alembert wave equation. 
Of course, this equation is 
conformally invariant, since if we choose null coordinates 
$z^+=t+z$ and $z^-=t-z$ one obtains 
\beq 
\partial_-\,\partial_+\,\phi=0\,, 
\label{dalemberteqnullcoordinates} 
\eeq 
and the change to coordinates ${z^+}'=f(z^+)$ and 
${z^-}'=f(z^-)$, which is a conformal transformation in 
two dimensions, yields the same wave equation 
(\ref{dalemberteqnullcoordinates}) 
for the scalar field. 
Moreover, using 
Eq. (\ref{f(z)}) when $z\rightarrow-\,\infty$, 
and Eq. (\ref{dalemberteq}), one finds that 
$T_{+-}=T_{-+}=0$, and thus 
the trace of $T_{ab}$ in coordinates $(z^+,z^-)$,  for large 
negative $z$, now obeys 
\beq 
T_{-+}+T_{+-}=0\,, 
\label{trace} 
\eeq 
i.e., it vanishes on-shell, confirming that the field $\phi$ is now 
described by a conformal field theory in an arbitrary small vicinity 
of the horizon \cite{Solodukhin}. Note that the conformal field 
theory is classical, since the trace $T^a_a$, given in 
(\ref{trace}), vanishes only when the equations of motion for 
$\phi$ are obeyed. 
Note also that on the horizon the field $\phi$ 
is only a function, either of the coordinate $z^+$, or of 
$z^-$. Since we are interested in the future branch of the 
horizon $\mathcal{H}_{\rm f}$ (see Sec. 
\ref{subsectionPreliminariesandmotivatiosection}), 
we take $\phi$ to be 
a function of  $z^+$. This will make 
the conformal field theory 
a chiral theory. 

\section{Conformal entropy} \label{conformalentropy} 

We are now ready to find the entropy, via conformal means, of these 
three classes of black holes, spherical, toroidal, and hyperbolic black 
holes. 

The conservation law for the energy-momentum tensor, 
$\partial_a T^{ab}=0$, 
in 
null coordinates is  $\partial_- T_{++}+\partial_+ T_{-+}=0$. 
As $T_{-+}=0$, this conservation 
law reads 
\beq 
\partial_- T_{++}=0\,. 
\label{conservedcharge0} 
\eeq 
Interchanging $-\leftrightarrow +$ one also finds, $\partial_+ 
T_{--}=0$.  Classically, incoming motion from the horizon is 
forbidden, so one should use only the component $T_{++}$ of the stress 
tensor as the physical charge. The corresponding charge generates the 
conformal transformations.  That is, if one has a function of $z^+$, 
e.g., $g(z^+)$, Eq. ({\ref{conservedcharge0}}) implies that 
$g\,T_{++}$ is conserved, $\partial_- (g\,T_{++})=0$. Then, the charge 
$Q_g=\int \,dz\, g(z^+)\,T_{++}$ is conserved, and there is an 
infinite set of them as well, for there is an infinite number of 
functions $g$.  This happens only in two dimensions.  The resemblance 
with bosonic string theory is now complete. 

The function $g(z^+)$ can be identified with $\xi^+(z^+)$, 
$g(z^+)\equiv\xi^+(z^+)$, where $\xi^+(z^+)$ is 
the non-zero component of  the 
generators of the conformal transformation 
$l_\xi=\xi^+\partial_+$ which preserve the 
horizon condition (\ref{hor}), 
(see subsection (\ref{subsectionPreliminariesandmotivatiosection})). 
Now, since we are 
interested in conserved quantities, these have the same value for any $t$, 
so we can calculate them at $t=0$, where the generators are only functions 
of 
$z$, i.e, 
$\xi^+\,\partial_+=\xi(z)\,\partial_z$, 
where we have also simplified the notation 
putting $\xi^+=\xi$. 
Thus the conserved charge 
associated with the infinitesimal conformal generators 
on the horizon $\xi\,\partial_z$ 
is 
then 
\beq 
Q_\xi=\int{ \,\xi(z)\,T_{++}}\,, 
\label{conservedcharge} 
\eeq 
where 
$T_{++}$ is evaluated at an infinitesimal vicinity 
of the horizon, also at $t=0$, such that 
$T_{++}=T_{++}(z)$, and can be found from Eq. 
(\ref{energymomentumtensor}), near the horizon. 
Since the Poisson algebra of theses charges, 
$\left\{Q_{\xi_1},Q_{\xi_2}\right\}$, is an expression of 
fundamental properties of the corresponding field theory, 
in this case a conformal one, 
one hopes that from this one obtains fundamental properties 
of the horizon. In order to obtain this algebra, note that 
from the action 
(\ref{actionfinalD>4}) one finds that the momentum $\pi$ conjugate to 
the field $\phi$ is $\pi=\delta\mathcal{L}/ \delta(\partial_t \phi) 
= \partial_t\phi$, 
where $\mathcal L$ is the Lagrangian associated to the action. 
Since the  Poisson algebra for 
the two fundamental fields $\phi$ and $\pi$ is defined as 
\cequ 
\anticomut{\phi(z,t)}{\pi(z',t)}&=&\int dy\left(\frac{\delta 
\phi(z,t)} 
{\delta \phi(y,t)}\frac{\delta \partial_t\phi(z',t)} 
{\delta \partial_t\phi(y,t)}-\frac{\delta \phi(z,t)}{\delta 
\partial_t\phi(y,t)} 
\frac{\delta \partial_t\phi(z',t)}{\delta \phi(y,t)}\right) \nonumber\\ 
&=&\delta(z-z')\,, 
\label{poissonalgebra} 
\fequ 
one finds, 
\beq 
\left\{Q_{\xi_1},Q_{\xi_2}\right\}=Q_{[\xi_1,\xi_2]}+ 
\left(\frac{q^2\phi_{\rm h}}{4}\right)^2 
\int{ C[\xi_1,\xi_2]}dz\,, 
\label{conservedchargecomm} 
\eeq 
where $[\xi_1,\xi_2]=\xi_1\xi'_2-\xi_2\xi'_1$, as before, 
with ${}^{'} \equiv \partial_z$, 
and $C[\xi_1,\xi_2]\equiv(\xi'_1+\beta_{\rm{h}}^{-1}\xi_1)(\xi'_1+ 
\beta_{\rm{h}}^{-1}\xi_2)'- 
(\xi'_1+\beta_{\rm{h}}^{-1}\xi_1)'(\xi'_1+\beta_{\rm{h}}^{-1}\xi_2)$. 
This is algebra is valid only in a close vicinity the horizon. 

Note that within the context 
of the pure conformal field theory (\ref{dalemberteq}), 
once it is granted that the field $\phi$ obeys a conformal 
field theory, 
the coordinate $z$ has a range $-\infty\leq z\leq+\infty$. 
In order to proceed, and be able to use standard techniques, 
one imposes periodic boundary conditions on the field, 
i.e., the field is confined within a box of length $L$, say, 
with an infinite number of boxes joined smoothly together 
so that the range of $z$ is still infinite. One can then concentrate 
in one box only of  length $L$. The 
length $L$ must have a physical meaning. In the context 
we are working with, the only quantum length scale that appears 
in the problem is the inverse Hawking temperature $\beta_{\rm h}$, 
which can be interpreted as the associated Compton thermal 
wavelength. As we will see this identification will appear 
naturally within this formalism. 
The best way to realize periodic boundary conditions 
here it is to compactify the coordinate $z$ 
on a circle of circumference $L$, such that $z$ 
lies in the interval 
$-\frac{L}{2}\leq z\leq\frac{L}{2}$. 

Then, the functions $\xi(z)$ 
can be expanded in Fourier modes as $\xi(z)= 
\sum \,a_n \,e^{i\frac{2\pi}{L}nz}$, 
where the $a_n$ 
are complex numbers. So, a basis for $\xi$ can be taken as 
$\xi_n=e^{i\frac{2\pi}{L}nz}$, i.e., we are looking 
at diffeomorphisms with period $L$. Then, the conserved charges 
$Q_{\xi}$ 
turn into conserved charges $Q_n$, given by 
\beq 
Q_{n}=\frac{L}{2\pi}\int_{-L/2}^{L/2}dz \,e^{i\frac{2\pi}{L}nz}\,T_{++}\,. 
\label{Qgenerators} 
\eeq 
These charges generate the following Poisson algebra 
\beq 
i\anticomut{Q_{k}}{Q_{n}}=(k-n)\,Q_{n+k}+\frac{c}{12} 
\left[k^{3}+\left(\frac{L}{2\pi\beta_{\rm{h}}}\right)^{2}\,k\right] 
\delta_{n+k,0}\;, 
\label{virasoro0algebra} 
\eeq 
which is a Virasoro type algebra, with 
\beq 
c=12\,\pi q^{4}\left(\frac{D-3}{D-2}\right)^2\phi^{2}_{\rm h}, 
\label{centralcharge} 
\eeq 
being the central charge. 

In order to transform (\ref{virasoro0algebra}) into 
the usual form of the Virasoro algebra one performs a 
shift from the generators $Q_n$  to generators $L_n$, 
given by 
\beq 
L_n=Q_n+\frac{c}{24}\,\left[\left(\frac{L}{2\pi\beta_{\rm{h}}}\right)^2 
+1\right]\,\delta_{n,0}\,. 
\label{shift0} 
\eeq 
The Virasoro generators $L_n$, which obey an algebra with a central 
charge, are prompt to be quantized. 
Note first that, near the horizon, 
all the functions depend only on the coordinate 
$z^+=t+z$, so that 
the periodicity properties in time $t$ should be identical 
to those in $z$. 
Then, since the most natural scale 
in the problem is the imaginary time 
period of the Euclidean black hole, one should set 
the length scale $L$ equal to it, 
$L=2\pi\,\beta_{\rm{h}}$. 
Thus, Eq. (\ref{shift0}) turns into 
\beq 
L_n=Q_n+\frac{c}{12}\,\delta_{n,0}\,. 
\label{shift} 
\eeq 
Second, to quantize the generators, one swaps the 
Poisson brackets for quantum commutators, 
through $[\;,\;]=i\hbar\{\;,\;\}$, and divides the generators 
by $\hbar$.  Then, with 
the help of Eq. (\ref{virasoro0algebra}) one finds 
that the conserved 
charge operators $L_n$,  usually 
called Virasoro operators, satisfy the usual Virasoro algebra, 
\beq 
[L_{k},L_{n}]=(k-n)\,L_{n+k}+\frac{c}{12} 
\left(k^3-k\right)\delta_{k+n,0}\,. 
\label{virasoroalgebra} 
\eeq 
Note that for $\phi_{\rm h}=0$, the case with no black hole, 
and thus no central charge, $c=0$, this set up 
gives the algebra of the conformal group on a classical 
massless scalar field $\phi$ in two dimensions, which is known 
to be a conformally invariant theory. 

Given that one has a Virasoro algebra for the conserved 
charge operators, one can now use the Cardy formula, 
which yields the number of asymptotic states 
of a conformal field theory. 
Its logarithm is the associated entropy, 
given by \cite{Cardy}, 
\beq 
S_{\rm conf}=2\pi\sqrt{ \frac{c}{6}\,\left( L_0-\frac{c}{24}\right)}\,\,, 
\label{Cardyformula} 
\eeq 
where $L_0\equiv {L_n}_{|_{n=0}}$ is now considered as the eigenvalue of 
the 
corresponding Virasoro operator. 

To find $L_0$ we first find $Q_0$ through Eq. 
(\ref{Qgenerators}).  Near the horizon $\phi\approx\phi_{\rm h}$, 
so in 
the semi-classical approximation, one has from 
(\ref{energymomentumtensor}) that $T_{++}$ is given by 
\cequ 
T_{++}=&&\frac{1}{4}(\partial_t\phi+\partial_z\phi)^2\nonumber 
\\ 
-&&\frac{1}{2}\, 
\left(\frac{D-3}{D-2}\right)\, 
q^2 \phi_{\rm h}\left[\partial_z(\partial_t\phi+\partial_z\phi)- 
\frac{2\,\pi}{\beta_{\rm{h}}} 
(\partial_t\phi+\partial_z\phi)\right]\,. 
\label{energymomentumtensoronhorizoninnullcoordinates} 
\fequ 
Then integration of $T_{++}$ along the coordinate $z$, 
compactified in a circle of 
circumference $L$, gives zero.  Since 
$Q_{0}=\frac{L}{2\pi}\int_{-L/2}^{L/2}dz \,T_{++}\,$, one has 
$Q_{0}=0$.  This is a consequence of the periodicity of the coordinate 
$z$ and of the fact that the integration is to be performed in the 
interval $-L/2\,<\,z\,<\,L/2$. Indeed, first note that 
$\int_{-L/2}^{L/2} dz \,\partial_z 
\phi=\left.\phi\right|_{-L/2}^{L/2}=0$, by periodicity. One also finds 
$\int_{-L/2}^{L/2} dz \,\partial_t \phi=\partial_t \, 
\int_{-L/2}^{L/2} dz \,\phi=0$ by periodicity. 
The same applies to the case where there are second 
derivatives, or squares of first derivatives.  For the case where we 
have mixed products, $\partial_t\phi \,\partial_z\phi$, we can see 
that $\partial_t\phi\, \partial_z\, 
\phi=\partial_t(\phi\,\partial_z\phi)-\phi\,\partial_t\partial_z\phi$. 
Using 
$\int_{-L/2}^{L/2} dz\,\phi\,\partial_z\phi=\frac{1}{2}\left. 
\phi^2\right|_{-L/2}^{L/2}=0$ and $\int_{-L/2}^{L/2} dz \, 
\partial_z\phi\,\partial_t\phi= 
\frac{1}{4}\left.\partial_t(\phi^2)\right|_{-L/2}^{L/2}=0$ we conclude 
that the charge $Q_0=0$. 
Then from Eq. (\ref{shift}) one finds 
that 
\beq 
L_0=\frac{c}{12}\,. 
\label{L0} 
\eeq 
Thus Solodukhin's method allows to obtain a relation between 
the eigenvalue $L_0$ and the central charge $c$, but not their values 
independently. Since the central charge depends on the 
unknown conformal parameter $q$, defined in Eqs. 
(\ref{redef1})-(\ref{redef3}), 
the  conformal entropy given by the 
Cardy formula (\ref{Cardyformula}) is determined 
up to the parameter $q$. Indeed, one has 
\beq 
S_{\rm{conf}}=2\,\pi^2 
q^4\left(\frac{D-3}{D-2}\right)^2\phi_{\rm h}^2\,. 
\label{conformalentropy1} 
\eeq 
Comparing the gravitational entropy given in 
the Eq. (\ref{entropiesingravitation1}), written in the form 
\beq 
S= \frac{A_{D-2}}{4\,G}= 
\frac{\Sigma_{D-2}^{\,,k} \,r_{\rm h}^{D-2}}{4\ G}=2\pi \,q^2 
\left(\frac{D-3}{D-2}\right)\phi^{2}_{\rm h}\,, 
\label{BHD} 
\eeq 
with the conformal entropy (\ref{conformalentropy1}), 
gives that 
\beq 
S_{\rm{conf}}=\left[\,q^2\, \left(\frac{D-3}{D-2}\right) \,\pi\, 
\right]\, S\,. 
\label{comparison} 
\eeq 
The parameter $q$ in Eqs. 
(\ref{redef1})-(\ref{redef3}) is a free parameter in the theory. 
It is a trade between fluctuations in the field 
$\phi$ and distance measurements through the metric $\bar{\gamma_{ab}}$. 
For small $q$ the fluctuations are large and the measurements 
small, yielding a large horizon entropy; 
for large $q$ the fluctuations are small and the measurements 
large, yielding a small entropy. If one chooses 
\beq 
q^2=\left(\frac{D-2}{D-3}\right)\frac{1}{\pi}\,, 
\label{q} 
\eeq 
one obtains precisely the Bekenstein-Hawking entropy. 

\section{Conclusions and remarks} \label{conclusions} 

In this analysis we have shown that the method of Solodukhin 
\cite{Solodukhin} may be applied to the more complete set of static 
black holes within general relativity, namely, black holes not only 
with a spherical horizon, but also with toroidal and hyperbolic 
compact horizons. We have shown that these additions, 
i.e., the introduction of different topologies for the horizon, were 
absorbed by the effective potential into the equations, which 
eventually disappear in the infinitesimal vicinity of the 
horizon. Thus, one obtains for the conformal entropy of the black hole 
that the Bekenstein-Hawking formula, i.e., the entropy is proportional 
to the horizon area, holds for any of the topologies considered within 
this formalism. This is another instance where the universality of 
the black hole entropy concept, as formulated in \cite{Carlip 
universality}, applies. 

On a more generic vein, one can remark that the horizon, through the 
boundary conditions imposed on it, gives a conformal character to the 
oscillatory dilaton field in its vicinity. This is no surprise, since 
it is known that, in general, the fields become massless 
asymptotically close to the horizon (see, e.g, 
\cite{PadmanabhanMPLA2002} for another instance of a scalar field 
propagation turning into a conformal field theory in the vicinity of 
the horizon).  What, perhaps, is surprising is that the corresponding 
classical conformal theory has a central charge, which upon 
quantization, and through appropriate methods gives, through Cardy 
formula, the black hole entropy as found in \cite{Solodukhin,Carlip 
6,Lin Wu,Cvitan et al Solodukhin} and here.  It is also worth noting 
that the Cardy formula applies to any conformal field theory in two 
dimensions, be it derived from condensed matter physics, quantum field 
theory, or gravity theories.  Explaining why very different theories, 
that nonetheless yield the same central charge and the same level 
operator $L_0$, have, through the Cardy formula, the same entropy, is 
no easy task, up to now there being no clear physical argument for 
that.  This is perhaps a consequence of the fact that the Cardy 
formula itself is hard to interpret 
\cite{CarlipCQGreviewofCardyformula}, an exception being the scalar 
field where a clear physical interpretation can be provided 
\cite{Fursaev}. 

\section*{Acknowledgments} 

This work was partially funded by Funda\c c\~ao para a Ci\^encia e a 
Tecnologia (FCT) of the Ministry of Science, Portugal, through project 
POCTI/FIS/57552/2004.  GASD is supported by grant SFRH/BD/2003 from 
FCT. JPSL thanks Observat\'orio Nacional do Rio de Janeiro for 
hospitality.

\end{document}